\documentclass[conference]{IEEEtran}
\IEEEoverridecommandlockouts
\usepackage{pgfplotstable}
\usepackage{pgfplots}
\pgfplotsset{
    /pgfplots/ybar legend/.style={
    /pgfplots/legend image code/.code={%
       \draw[##1,/tikz/.cd,yshift=-0.25em]
        (0cm,0cm) rectangle (3pt,0.8em);},
   },
}
\usepackage{url}
\usetikzlibrary{patterns}
\usepackage{graphicx}
\usepackage{listings}
\usepackage{courier}
\usepackage{balance}
\usepackage{subcaption}
    \newenvironment{customlegend}[1][]{%
        \begingroup
        \csname pgfplots@init@cleared@structures\endcsname
        \pgfplotsset{#1}%
    }{%
        \csname pgfplots@createlegend\endcsname
        \endgroup
    }%

    \def\addlegendimage{\csname pgfplots@addlegendimage\endcsname}

\usepackage{xcolor,colortbl}
\usepackage{booktabs}
\usepackage{hyperref}
\usepackage{tabulary}
\usepackage{color}
\title{Mining Sandboxes for Linux Containers}
\author{\IEEEauthorblockN{Zhiyuan Wan\IEEEauthorrefmark{1}, David Lo\IEEEauthorrefmark{2}, Xin Xia\IEEEauthorrefmark{1}$^\ddag$\thanks{$^\ddag$Corresponding author}, Liang Cai\IEEEauthorrefmark{1}, and Shanping Li\IEEEauthorrefmark{1}}
 \IEEEauthorblockA{\IEEEauthorrefmark{1}College of Computer Science and Technology, Zhejiang University, China}
 \IEEEauthorblockA{\IEEEauthorrefmark{2}School of Information Systems, Singapore Management University, Singapore
 \\ \{wanzhiyuan, xxkidd, leoncai, shan\}@zju.edu.cn, davidlo@smu.edu.sg
   }}
\begin{document}
\maketitle
\begin{abstract}
A container is a group of processes isolated from other groups via distinct kernel namespaces and resource allocation quota. Attacks against containers often leverage kernel exploits through system call interface. In this paper, we present an approach that mines sandboxes for containers. We first explore the behaviors of a container by leveraging automatic testing, and extract the set of system calls accessed during testing. The set of system calls then results as a sandbox of the container. The mined sandbox restricts the container's access to system calls which are not seen during testing and thus reduces the attack surface. In the experiment, our approach requires less than eleven minutes to mine sandbox for each of the containers. The enforcement of mined sandboxes does not impact the regular functionality of a container and incurs low performance overhead.
\end{abstract}
\section{Introduction}

Platform-as-a-Service (PaaS) cloud  is a fast-growing segment of  cloud market, being projected to reach \$7.5 billion by 2020 \cite{paas}. A PaaS cloud permits tenants to deploy applications in the form of application executables or interpreted source code (e.g. PHP, Ruby, Node.js, Java). The deployed applications execute in a provider-managed host OS, which is shared with applications of other tenants. Thus a PaaS cloud often leverages OS-based techniques, such as Linux containers, to isolate applications and tenants.


Containers provide a lightweight operating system level virtualization, which groups resources like processes, files and devices into isolated namespaces. This gives users the appearance of having their own operating system with near native performance and no additional virtualization overhead. Container technologies, such as Docker \cite{merkel2014docker}, enable an easy packaging and rapid deployment of applications. However, most containers that run in the cloud are too complicated to trust. The primary source of security problems in containers is system calls that are not namespace-aware \cite{felter2015updated}. Non-namespace-aware system call interface facilitates the adversary to compromise untrusted containers to exploit kernel vulnerabilities to elevate privileges, bypass access control policy enforcement, and escape isolation mechanisms. For instance, a compromised container can exploit a bug in the underlying kernel that allows privilege escalation and arbitrary code execution on the host \cite{cve}.

How can cloud providers protect the clouds from untrusted containers? One straightforward way is to place the container in a sandbox to restrain its access to system calls. By restricting system calls, we could also limit the impact that an adversary can make if a container is compromised. System call interposition is a powerful approach to restrict the power of a program by intercepting its system calls \cite{garfinkel2003traps}. Sandboxing techniques based on system call interposition have been developed in the past \cite{goldberg1996secure,provos2003improving,acharya2000mapbox,fraser1999hardening,ko2000detecting,kim2013practical}. Most of them focus on implementing sandboxing techniques and ensuring secure system call interposition. However, generating accurate sandbox policies for a program are always challenging \cite{provos2003improving}. We are inspired by a recent work \emph{BOXMATE} \cite{jamrozik2016mining}, which learns and enforces sandbox policies for Android applications. \emph{BOXMATE} first explores Android application behavior and extracts the set of resources accessed during testing. This set is then used as a sandbox, which blocks access to resources not used during testing. We would like to port the idea of \emph{sandbox mining} in \emph{BOXMATE} to be able to confine Linux containers.


\begin{figure}
\includegraphics[width=\linewidth]{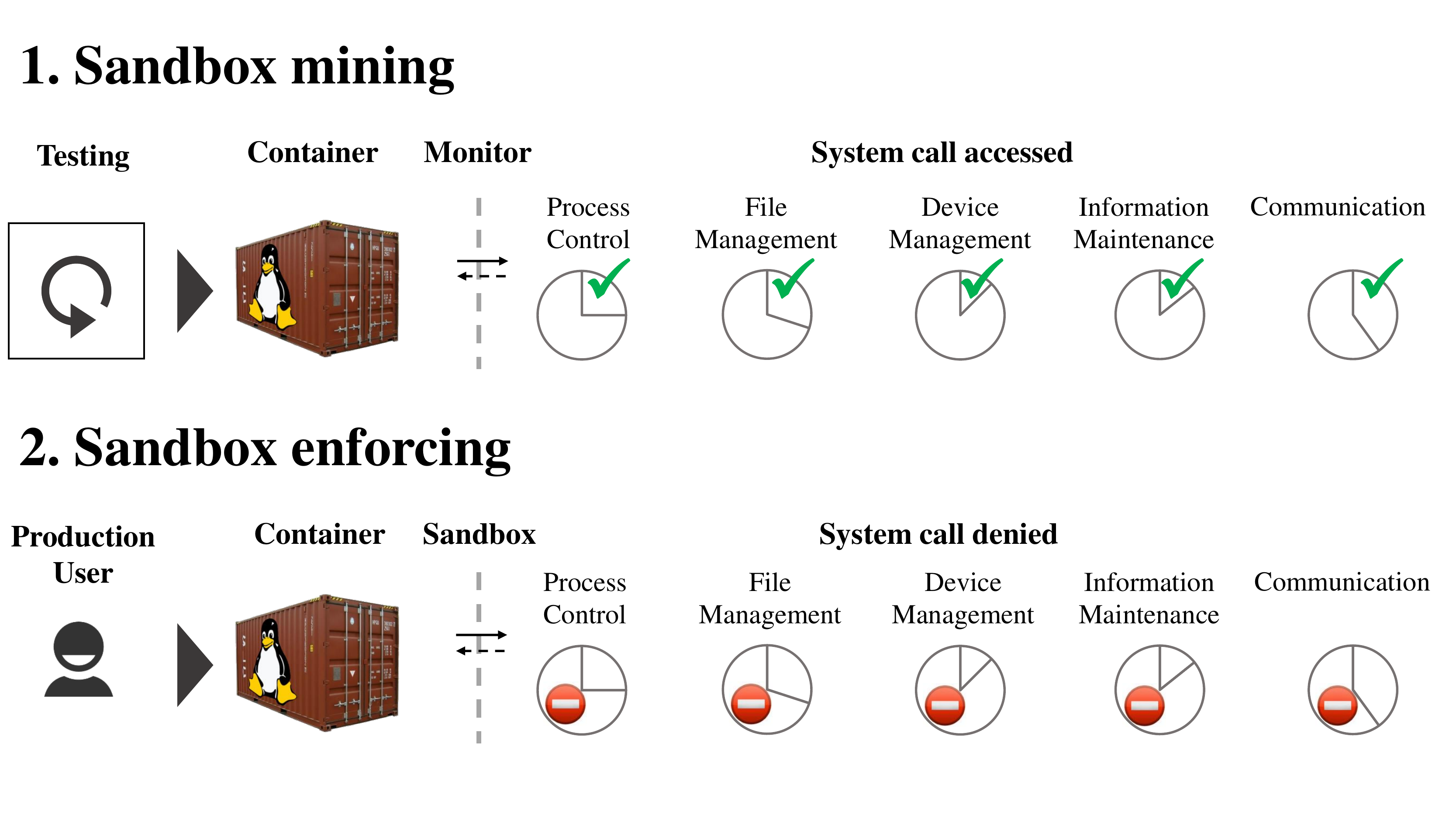}
\caption{Our approach in a nutshell. Mining phase monitors accessed system calls when testing. These system calls make up a \emph{sandbox} for the container, which later prohibits access to system calls not accessed during testing. }\vspace{-0.5cm}
\label{fig:phases}
\end{figure}
A container comprises multiple processes of different functionalities that access distinct system calls. Different containers may access distinct sets of system calls. Therefore, a common sandbox for all the containers is too coarse. In this paper, we present an approach to \emph{automatically extract sandbox rules} for a given container. The approach is composed of two phases shown in Fig. \ref{fig:phases}:
\begin{itemize}
\item
\textbf{Sandbox mining.} In the first phase, we mine the rules that will make the sandbox. We use automatic testing to explore container behaviors, and monitor all accesses to system calls.
\item
\textbf{Sandbox enforcing.} In the second phase, we assume that system calls which are not accessed during the mining phase should not be accessed in production either. Consequently, if the container (unexpectedly) requires access to a new system call, the sandbox will prohibit access.
\end{itemize}

To the best of our knowledge, our approach is the first technique to leverage automatic testing to extract sandbox rules for Linux containers. While our approach is applicable to any Linux container management service, we selected Docker as a concrete example because of its popularity. Our approach has a number of compelling features:
\begin{itemize}
\item \textbf{No training in production.} In contrast to anomaly detection systems, our approach does not require training process in production. The ``normal'' system call access would already be explored during the mining phase.
\item \textbf{Reducing attack surface. } The mined sandbox detects system calls that cannot be seen during the mining phase, which reduces the attack surface by confining the adversary and limiting the damage he/she could cause.
\item \textbf{Guarantees from sandboxing. } Our approach runs test suites to explore ``normal'' container behaviors. The testing may be incomplete, and other (in particular malicious) behaviors are still possible. However, the testing covers a safe subset of all possible container behaviors. Sandboxing is then used to guarantee that no unknown system calls aside from those used in the testing phase are permitted.
\end{itemize}

We evaluate our approach by applying it to eight Docker containers and focus on three research questions:

\noindent \textbf{RQ1. How efficiently can our approach mine sandboxes?}

We automatically run test suites on Docker containers, and check the system call convergence. It takes less than two minutes for the set of accessed system calls to saturate. In addition, we compare our mined sandboxes with the default sandbox provided by Docker. The default sandbox allows more than 300 system calls \cite{dockerseccomp} and is thus too coarse. On the contrary, our mined sandboxes allow 66 - 105 system calls for eight containers in the experiment, which significantly reduce the attack surface.

\noindent \textbf{RQ2. Can automatic testing sufficiently cover behaviors?}

If a system call $S$ is not accessed during the mining phase, later non-malicious access to $S$ would trigger a false alarm.  We run use cases that cover core functionality of containers to check whether the enforcing mined sandboxes would trigger alarms. The result shows that all the use cases end with no false alarms.

\noindent \textbf{RQ3. What is the performance overhead of sandbox enforcement?}

We evaluate the performance overhead of enforcing mined sandboxes on a set of containers. The result shows that sandbox enforcement incurs very low end-to-end performance overhead (0.6\% - 2.14\%). Our mined sandboxes also provide a slightly lower performance overhead than that of the default sandbox.

The remainder of this paper is organized as follows. After discussing background and related work in Section \ref{sec:backgroundrelatedwork}, Section \ref{sec:motivation} specifies the threat model and motivation of our work. Section \ref{sec:mine} and \ref{sec:enforce} detail two phases of our approach. We evaluate our approach in Section \ref{sec:experiment} and discuss threats to validity and limitations in Section \ref{sec:threat}. Finally, Section \ref{sec:conclusion} closes with conclusion and future work.
\section{Background and Related Work}
\label{sec:backgroundrelatedwork}
\subsection{System Call Interposition}

System calls allow virtually all of a program's interactions with the network, filesystem, and other sensitive system resources. System call interposition is a powerful approach to restrict the power of a program~\cite{garfinkel2003traps}.

There exists a significant body of related work in the domain of system call interposition. Implementing system call interposition tools securely can be quite subtle \cite{garfinkel2003traps}. Garfinkel studies the common mistakes and pitfalls, and uses the system call interposition technique to enforce security policies in the Ostia tool \cite{garfinkel2004ostia}. System call interposition tools, such as Janus \cite{goldberg1996secure,wagner1999janus}, Systrace \cite{provos2003improving}, and ETrace \cite{jain2000user}, can enforce fine-grained policies at granularity of the operating system's system call infrastructure. System call interposition is also used for sandboxing \cite{goldberg1996secure,provos2003improving,acharya2000mapbox,fraser1999hardening,ko2000detecting,kim2013practical} and intrusion detection \cite{hofmeyr1998intrusion,forrest1996sense,wagner2001intrusion,bhatkar2006dataflow,kiriansky2002secure,warrender1999detecting,somayaji2000automated,sekar2001fast,mutz2006anomalous}.

\emph{Seccomp-BPF} framework \cite{seccompbpf} is a system call interposition implementation for Linux Kernel introduced in Linux 3.5. It is an extension to \emph{Seccomp} \cite{seccomp}, which is a mechanism to isolate a third-party application by disallowing all system calls except for reading and writing of already-opened files. \emph{Seccomp-BPF} generalizes \emph{Seccomp} by accepting \emph{Berkeley Packet Filter} (BPF) programs to filter system calls and their arguments. For example, the \emph{BPF} program can decide whether a program can invoke the {\tt reboot()} system call.

In Docker, the host can assign a \emph{Seccomp BPF} program for a container. Docker uses a \emph{Seccomp profile} to capture a \emph{BPF} program for readability \cite{dockerseccomp}. Fig. \ref{fig:seccomp_profile_sample} shows a snippet of \emph{Seccomp profile} used by Docker, written in the JSON \cite{json} format.

\begin{figure}[btp]
\begin{lstlisting}
{
	"defaultAction": "SCMP_ACT_ERRNO",
	"architectures": [
		"SCMP_ARCH_X86_64",
		"SCMP_ARCH_X86",
		"SCMP_ARCH_X32"
	],
	"syscalls": [
		{
			"name": "accept",
			"action": "SCMP_ACT_ALLOW",
			"args": []
		},
		{
			"name": "accept4",
			"action": "SCMP_ACT_ALLOW",
			"args": []
		},
		...
	]
}
\end{lstlisting}
\caption{A snippet of Docker \emph{Seccomp profile}, expressed in JavaScript Object Notation (JSON).}
\label{fig:seccomp_profile_sample}\vspace{-0.5cm}
\end{figure}

By default, Docker disallows 44 system calls out of 300+ for all of the containers to provide wide application compatibility \cite{dockerseccomp}. However, the principle of least privilege \cite{saltzer1975protection} requires that a program must only access the information and resources necessary to complete its operation. In our experiment, we notice that top-downloaded Docker containers access less than 34\% of the system calls which are whitelisted in the default \emph{Seccomp profile}.

\vspace{0.1cm}\noindent\fbox{%
    \parbox{0.965\linewidth} {%
      \emph{
        Containers are granted more privileges than they require.
        }
    }
}

\subsection{System Call Policy Generation}
Generating an accurate system call policy for an existing program has always been challenging \cite{provos2003improving}. It is difficult and impossible to generate an accurate policy without knowing all possible behaviors of a program. The question ``what does a program do?'' is the general problem of \emph{program analysis}. Program analysis falls into two categories: \emph{static} analysis and \emph{dynamic} analysis.

Static analysis checks the code without actually executing programs. It sets an upper bound to what a program can do. If static analysis determines some behavior is impossible, the behavior can be safely excluded. Janus \cite{goldberg1996secure} recognizes a list of dangerous system calls statically. Wagner and Dean \cite{wagner2001intrusion} derive system call sequences from program source code.

The limitation of static analysis is \emph{over-approximation}. The analysis often assume that more behaviors are possible than actually would be. Static analysis is also undecidable in all generality due to the halting problem.

\vspace{0.1cm}\noindent\fbox{%
    \parbox{0.965\linewidth} {%
         \emph{
        Static analysis produces over-approximation.
        }
    }%
}\vspace{0.2cm}

Dynamic analysis analyzes actual executions of a running program. It sets a lower bound of a program's behaviors. Any (benign) behavior seen in past executions should be allowed in the future as well. Given a set of executions, one can learn program benign behaviors to infer system call policies. There is a rich set of articles about system call policy generation through dynamic analysis. Some studies look at a sequence of system calls to detect deviations to normal behaviors~\cite{forrest1996sense,hofmeyr1998intrusion,somayaji2000automated}. Instead of analyzing system call sequences, some studies take into account the arguments of system calls. \cite{sekar2001fast} uses finite state automata (FSA) techniques to capture temporal relationships among system calls~\cite{mutz2006anomalous,kruegel2003detection}. Some studies keep track of data flow between system calls~\cite{bhatkar2006dataflow,fetzer2008switchblade}. Other researchers also take advantage of machine learning techniques, such as Hidden Markov Models (HMM)  \cite{warrender1999detecting,gao2006behavioral}, Neural Networks \cite{endler1998intrusion}, and k-Nearest Neighbors \cite{liao2002use}.

The fundamental limitation of dynamic analysis is \emph{incompleteness}. If some behavior has not been observed so far, there is no guarantee that it may not occur in the future. Given the high cost of false alarms, a sufficient set of executions must be available to cover all of the normal behaviors. The set of executions can either derive from testing, or from production (a training phase is required) \cite{jamrozik2016mining}.

\vspace{0.1cm}\noindent\fbox{%
    \parbox{0.965\linewidth} {%
         \emph{
        Dynamic analysis requires sufficient ``normal'' executions to be trained with.
        }
    }%
}

\subsection{Consequences}
Sandboxing, program analysis and testing are mature technologies. However, each of them has limitations: sandboxing needs policy, dynamic analysis needs executions, and testing cannot guarantee the absence of malicious behavior \cite{jamrozik2016mining}. Nonetheless,  Zeller et al. argue that combining the three not only mitigates the limitations, but also \emph{turns the incompleteness of dynamic analysis into a guarantee}~\cite{zeller2015test}. In our case, system call interposition-based sandboxing can guarantee that anything not seen yet will not happen.

\section{Threat Model and Motivation}
\label{sec:motivation}
Most containers that run in the cloud, e.g., Web server, database systems and customized applications, are too complicated to trust. Even with access to the source code, it is difficult to reason about the security of a container. An untrusted container might be compromised by carefully craft inputs because of exploitable vulnerabilities. A compromised container can further do harm in many ways. For instance, a compromised container can exploit a bug in the underlying kernel that allows privilege escalation and arbitrary code execution on the host \cite{cve}; it can also acquire packet of another container via ARP spoofing \cite{whalen2001introduction}. We assume the existence of vulnerabilities to the adversary that he/she can use to gain unauthorized access to the underlying operating system and further compromise other containers in the cloud.

We observe that system call interface is the only gateway to make persistent changes to the underlying systems \cite{provos2003improving}. Nevertheless, system call interface is dangerously wide; less-exercised system calls are a major source of kernel exploits. To limit the impact an adversary can make, it is straightforward to sandbox a container and restrict the system calls it is permitted to access. We notice that the default sandbox provided by Docker disallows only 44 system calls -- the default sandbox is too coarse. Containers are granted more privileges than they require. To follow the
principle of least privilege, our approach automatically mines sandbox rules for containers during testing; and later enforces the policy by restricting system call invocations through sandboxing.
\section{Sandbox Mining}
\label{sec:mine}
\subsection{Overview}
During the mining phase, we automatically explore container behaviors, and monitor its system calls. This section illustrates the fundamental steps of our approach during the mining phase.
\subsubsection{\textbf{Enable tracing}} The first step is to prepare the kernel to enable tracing. We use container-aware monitoring tool \emph{sysdig} \cite{sysdig} to record system calls that are accessed by a container at run time. The monitoring tool \emph{sysdig} logs:
\begin{itemize}
\item an \emph{enter} entry for a system call, including timestamp, process that executes the system call, thread ID (which corresponds to the process ID for single-threaded processes), and list of system call arguments;
\item an \emph{exit} entry for a system call, with the properties mentioned above, except that replacing the list of arguments with return value of the system call.
\end{itemize}

\subsubsection{\textbf{Automatic testing}}
In this step, we select a test suite that covers functionality of a container. Then we run the test suite on the targeted container. During testing, we automatically copy the tracing logs at constant time intervals. This allows us to compare at what time system call was accessed. Therefore, we can monitor the growth of the sandbox rules overtime based on these snapshots.

\subsubsection{\textbf{Extract system calls}}
A script extracts the set of system calls accessed by a container from the tracing logs.
\begin{figure}[btp]
\begin{lstlisting}
[github.com/opencontainers/runc/libcontainer/utils/utils_unix.go: CloseExecFrom]
(*\bfseries 1  openat()*)
(*\bfseries 2  getdents64()*)
(*\bfseries 3  lstat()*)
(*\bfseries 4  close()*)
(*\bfseries 5  fcntl()*)
[github.com/opencontainers/runc/libcontainer/capabilities_linux.go: newCapWhitelist]
(*\bfseries 6  getpid()*)
(*\bfseries 7  capget()*)
[github.com/opencontainers/runc/libcontainer/system/linux.go: SetKeepCaps]
(*\bfseries 8  prctl()*)
[github.com/opencontainers/runc/libcontainer/init_linux.go: setupUser]
(*\bfseries 9  getuid()*)
(*\bfseries 10 getgid()*)
(*\bfseries 11 read()*)
[github.com/opencontainers/runc/libcontainer/init_linux.go: fixStdioPermissions]
(*\bfseries 12 stat()*)
(*\bfseries 13 fstat()*)
(*\bfseries 14 fchown()*)
[github.com/opencontainers/runc/libcontainer/init_linux.go: setupUser]
(*\bfseries 15 setgroups()*)
[github.com/opencontainers/runc/libcontainer/system/syscall_linux_64.go: Segid]
(*\bfseries 16 setgid()*)
[github.com/opencontainers/runc/libcontainer/system/syscall_linux_64.go: Seuid]
(*\bfseries 17 futex()*)
(*\bfseries 18 setuid()*)
[github.com/opencontainers/runc/libcontainer/capabilities_linux.go: drop]
(*\bfseries 19 capset()*)
[github.com/opencontainers/runc/libcontainer/init_linux.go: finalizeNamespace]
(*\bfseries 20 chdir()*)
[github.com/opencontainers/runc/libcontainer/standard_init_linux.go: Init]
(*\bfseries 21 getppid()*)
[github.com/opencontainers/runc/libcontainer/system/linux.go: Execv]
(*\bfseries 22 execve()*)
[github.com/docker-library/hello-world/hello.c: _start()]
(*\bfseries 23 write()*)
(*\bfseries 24 exit()*)
\end{lstlisting}
\caption{24 system calls accessed by \emph{hello-world} container discovered by our approach, and functions (in {\tt[]}) that first trigger them.}\vspace{-0.5cm}
\label{fig:hello_world_syscall_list}
\end{figure}

\subsection{Case Study}
As an example of how our approach explores container behaviors, let us consider \emph{hello-world} container \cite{helloworld}.
This container employs a Docker image which simply prints out a message and does not accept inputs. We discover 24 system calls during testing. The actual system calls are listed in Fig. \ref{fig:hello_world_syscall_list}. Docker {\tt init} process \cite{runc_standard_init_linux_v011} and \emph{hello-world} container invoke the system calls as follows:
\begin{itemize}
\item{\textbf{SYSCALL 1}} Right after the \emph{Seccomp profile} is applied, Docker {\tt init} process closes all unnecessary file descriptors that are accidentally inherited by accessing {\tt openat()} , {\tt getdents64()}, {\tt lstat()}, {\tt close()}, and {\tt fcntl()}.
\item {\textbf{SYSCALL 6}}
    Then Docker {\tt init} process creates a whitelist of capabilities with the process information by accessing {\tt getpid()} and {\tt capget()}.
\item {\textbf{SYSCALL 8}}
    Docker {\tt init} process preserves the existing capabilities by accessing {\tt prctl()} before changing user of the process.
\item {\textbf{SYSCALL 9}}
    Docker {\tt init} process obtains the user ID and group ID by accessing {\tt getuid()} and {\tt getgid()}; Later it reads the groups and password information from configuration file by accessing {\tt read()}.
\item {\textbf{SYSCALL  12}}
    Docker {\tt init} process fixes the permissions of standard I/O file descriptors by accessing {\tt stat()}, {\tt fstat()}, and {\tt fchown()}. Since these file descriptors are created outside of the container, their ownership should be fixed and match the one inside the container.
\item {\textbf{SYSCALL 15}}
    Docker {\tt init} process changes groups, group ID, and user ID for current process by accessing {\tt setgroups()}, {\tt setgid()}, {\tt futex()} and {\tt setuid()}.
\item {\textbf{SYSCALL 19}}
    Docker {\tt init} process drops all capabilities for current process except those specified in the whitelist by accessing {\tt capset()}.
\item {\textbf{SYSCALL 20}}
    Docker {\tt init} process changes current working directory to the one specified in the configuration file by accessing {\tt chdir()}.
\item {\textbf{SYSCALL 21}}
    Docker {\tt init} process then compares the parent process with the one from the start by accessing {\tt getppid()} to make sure that the parent process is still alive.
\item {\textbf{SYSCALL 22}}
    The final step of Docker {\tt init} process is accessing {\tt execve()} to execute the initial command of \emph{hello-world} container.
\item {\textbf{SYSCALL 23}}
    The initial command of \emph{hello-world} container executes {\tt hello} program. The {\tt hello} program writes a message to standard output (file descriptor 1) by accessing {\tt write()} and finally exits by accessing {\tt exit()}.
\end{itemize}
Ideally, we expect to capture the set of system calls accessed only by the container. However, the captured set include some system calls that are accessed by Docker {\tt init} process. This is because applying sandbox rules is a privileged operation; Docker {\tt init} process should apply sandbox rules before dropping capabilities. We notice that Docker {\tt init} process invokes 22 system calls to prepare runtime environment before the container starts. If Docker {\tt init} process accesses fewer system calls before the container starts, our mined sandboxes could be more fine-grained.

The system calls characterize the resources that \emph{hello-world} container accesses in our run. Since the container does not accept any inputs, we find the 24 system calls are an exhausted list. The testing will be more complicated if a container accepts inputs to determine its behavior.

\section{Sandbox Enforcing}
\label{sec:enforce}
\subsection{Overview}
The second phase of our approach is sandbox enforcing, which monitors and possibly prevents container behavior. We need a technique that conveniently allows user to sandbox any container. To this end, we leverage \emph{Seccomp-BPF} \cite{seccompbpf} for sandbox policy enforcement. Docker uses operating system virtualization techniques, such as \emph{namespaces}, for container-based privilege separation. \emph{Seccomp-BPF} further establishes a restricted environment for containers, where more fine-grained security policy enforcement takes place. During sandbox enforcement, the applied \emph{BPF} program checks whether an accessed system call is allowed by corresponding sandbox rules. If not, the system call will return an error number; or the process which invokes that system call will be killed; or a \emph{ptrace} event \cite{ptrace} is generated and sent to the tracer if there exists one. This section illustrates the two steps of our approach during sandboxing phase.

\subsubsection{\textbf{Generate sandbox rules}}

This step translates the set of system calls discovered in mining phase into sandbox rules using \emph{awk} tool. For instance, {\tt write()} is one of the discovered system calls during sandbox mining for \emph{hello-world} container. It will be translated to a sandbox rule with name {\tt write}, action {\tt SCMP\_ACT\_ALLOW}, and no constraint applied to the arguments ({\tt args}) as follows:
\begin{lstlisting}
{
    "name": "write",
	"action": "SCMP_ACT_ALLOW",
	"args": []
}
\end{lstlisting}
When the system call {\tt write()} is accessed during sandboxing, it will be permitted according to the specified action, i.e., {\tt SCMP\_ACT\_ALLOW}. After translating each system call entry into a sandbox rule, these rules constitute a whitelist of system calls that are allowed by the sandbox. We define the default action of the sandbox as follows:
\begin{lstlisting}
"defaultAction": "SCMP_ACT_ERRNO"
\end{lstlisting}
During sandboxing, when a container accesses a system call which is not included in the whitelist, the sandbox will deny the system call and make the system call return an error number ({\tt SCMP\_ACT\_ERRNO}).

\subsubsection{\textbf{Enforce sandbox rules}}
The resulting \emph{Seccomp profile} now contains all sandbox rules that allow the system calls observed in the mining phase. We then start the container with the \emph{Seccomp profile} to enforce mined sandbox rules using {\tt docker run --security-opt seccomp}.
\subsection{Case Study}
As an example of how our approach operates, consider \emph{hello-world} container again. The default Docker sandbox allows more than 300 system calls, which is a considerable attack surface. In that default setting, a compromised \emph{hello-world} container could simply mount a directory that contains a carefully crafted program. The program could open a socket by accessing system call {\tt socket()}, which is an abnormal behavior. By enforcing our mined sandbox, \emph{hello-world} container is not allowed to access {\tt socket()}. Thus we prevent the container from opening a socket and doing further harm.

\section{Experiments}
\label{sec:experiment}
\subsection{Overview}
In this section, we evaluate our approach to answer three research questions as follows:

\noindent \textbf{RQ1. How efficiently can our approach mine sandboxes?}

We evaluate how fast the sets of system calls are saturated  for eight containers. Notice that the eight containers are the most popular containers in Docker Hub \cite{dockerhub} and have a large number of downloads. The details of them are shown in TABLE~\ref{tab:subjects}.  The eight containers can be used in PaaS, and provide domain-specific functions rather than basic functions provided by OS containers (e.g. \emph{Ubuntu} container).
Note that \emph{python} as a programming language provides a wide range of functionality, and \emph{python} container can potentially access all system calls. Mining sandbox for \emph{python} container will be useless because the mined sandbox will be too coarse. Thus we setup Web framework \emph{Django} \cite{django} on top of \emph{python} container. This makes \emph{python} container have specific functionality. In addition, we compare the mined sandboxes with the default one provided by Docker to see if the attack surface is reduced.

\begin{table*}[t]
  \centering
  \caption{Experiment subjects. Open \url{https://hub.docker.com/_/}$<$identifier$>$ for details.}
    \begin{tabular} 
    {l r l r r l}

    \rowcolor[rgb]{ .651,  .651,  .651} \textbf{Name} & \textbf{Version} & \textbf{Description} & \textbf{Stars} & \textbf{Pulls} & \textbf{Identifier (links to Web page)} \\
    Nginx & 1.11.1 & Web server & 3.8K  & 10M+  & nginx \\
    \rowcolor[rgb]{ .851,  .851,  .851} Redis & 3.2.3 & key-value database & 2.5K  & 10M+  & redis \\
    MongoDB & 3.2.8 & document-oriented database & 2.2K  & 10M+  & mongo \\
    \rowcolor[rgb]{ .851,  .851,  .851} MySQL & 5.7.13 & relational database & 2.9K  & 10M+  & mysql \\
    PostgreSQL & 9.5.4 & object-relational database & 2.5K  & 10M+  & postgres \\
    \rowcolor[rgb]{ .851,  .851,  .851} Node.js & 6.3.1 & Web server & 2.6K  & 10M+  & node \\
    Apache & 2.4.23 & Web server & 606   & 10M+  & httpd \\
    \rowcolor[rgb]{ .851,  .851,  .851} Python & 3.5.2 & programming language & 1.1K  & 5M+   & python \\
    \end{tabular}\vspace{-0.5cm}%
  \label{tab:subjects}%
\end{table*}%

\noindent \textbf{RQ2. Can automatic testing sufficiently cover behaviors?}

Any non-malicious system call behavior not explored during testing implies a false alarm during production. We evaluate the risk of false alarms: how likely is it that sandbox mining misses functionality, and how frequently will containers encounter false alarms. We check the mined sandboxes of the eight containers against the use cases. We carefully read the documentation of the containers to make sure the use cases reflect the containers' typical usage.

\noindent \textbf{RQ3. What is the performance overhead of sandbox enforcement?}

As a security mechanism, the performance overhead of sandbox enforcement should be small. Instead of CPU time, we measure the end-to-end performance of containers -- \emph{transactions per second}.  We compare the end-to-end performance of a container running in three environments: 1) natively without sandbox, 2) with a sandbox mined by our approach, and 3) with default Docker sandbox.

The containers in the experiments run on a 64-bit Ubuntu 16.04 operating system inside VirtualBox 5.0.24 (4GB base memory, 2 processors). The physical machine is with an Intel Core i5-6300 processor and 8GB memory.

\subsubsection{Testing}
We describe the test suites we run in the experiment as follows:
\leavevmode \newline
\textbf{Web server (Nginx, Apache, Node.js, and Python Django).} After executing {\tt docker run}, each container experiences a warm-up phase which lasts for 30 seconds. After the warm-up phase, the Web server gets ready to serve requests. We remotely start with a simple HTTP request using \emph{wget} tool from another virtual machine. The request fetches a file from the server right after the warm-up phase. It is followed by a number of runs of \emph{httperf} tool \cite{mosberger1998httperf} also from that virtual machine. \emph{httperf} continuously accesses the static pages hosted by the container. The workload starts from 5 requests per second, increases the number of requests by 5 for every run, and ends at 50 requests per second.

\noindent\textbf{Redis.}
The warm-up phase of \emph{Redis} container lasts for 30 seconds. After the warm-up phase, we locally connect to the \emph{Redis} container via {\tt docker exec}. Then we run the built-in benchmark test \emph{redis-benchmark} \cite{redisbench} with the default configuration, i.e., 50 parallel connections, totally 100,000 requests, 2 bytes of SET/GET value, and no pipeline. The test cases cover the commands as follows:
\begin{itemize}
\item \textbf{PING}: checks the bandwidth and latency.
\item \textbf{MSET}: replaces multiple existing values with new values.
\item \textbf{SET}: sets a key to hold the string value.
\item \textbf{GET}: gets the value of some key.
\item \textbf{INCR}: increments the number stored at some key by one.
\item \textbf{LPUSH}: inserts all the specified values at the head of the list.
\item \textbf{LPOP}: removes and returns the first element of the list.
\item \textbf{SADD}: adds the specified members to the set stored at some key.
\item \textbf{SPOP}: removes and returns one or more random elements from the set value.
\item \textbf{LRANGE}: returns the specified elements of the list.
\end{itemize}
\textbf{MongoDB.}
The warm-up phase of \emph{MongoDB} container lasts for 30 seconds. After the warm-up phase, we run \emph{mongo-perf} \cite{mongoperf} tool to connect to \emph{MongoDB} container remotely from another virtual machine. \emph{mongo-perf} measures the throughput of \emph{MongoDB} server. We run each of the test cases in \emph{mongo-perf} with tag \emph{core}, on 1 thread, and for 10 seconds. The detail of test cases is described as follows:
\begin{itemize}
\item \textbf{insert document}: inserts documents only with object ID into collections.
\item \textbf{update document}: randomly selects a document using object ID and increments one of its integer field.
\item \textbf{query document}: queries for a random document in the collections based on an indexed integer field.
\item \textbf{remove document}: removes a random document using object ID from the collections.
\item \textbf{text query}: runs case-insensitive single-word text query against the collections.
\item \textbf{geo query}: runs \emph{nearSphere} query with \emph{geoJSON} format and two-dimensional sphere index.
\end{itemize}
\textbf{MySQL.}
The warm-up phase of \emph{MySQL} container lasts for 30 seconds. After the warm-up phase, we create a database, and use \emph{sysbench} \cite{sysbench} tool to connect to \emph{MySQL} container. We then run the \emph{OLTP} database test cases in \emph{sysbench} with maximum request number of 800, on 8 threads for 60 seconds. The test cases include the following functionalities:
\begin{itemize}
\item \textbf{create database}: creates a database \emph{test}.
\item \textbf{create table}: creates a table \emph{sbtest} in the database.
\item \textbf{insert record}: inserts 1,000,000 records into the table.
\item \textbf{update record}: updates records on indexed and non-indexed columns.
\item \textbf{select record}: selects records with a record ID and a range for record ID.
\item \textbf{delete records}: deletes records with a record ID.
\end{itemize}
\textbf{PostgreSQL.}
The warm-up phase of \emph{PostgreSQL} container lasts for 30 seconds. After the warm-up phase, we connect to \emph{PostgreSQL} container using \emph{pgbench} \cite{pgbench} tool. We first run \emph{pgbench} initialization mode to prepare the data for testing. The initialization is followed by two 60-second runs of read/write test cases with queries. The test cases cover the functionalities as follows:
\begin{itemize}
\item \textbf{create database}: creates a database {\tt pgbench}.
\item \textbf{create table}: creates four tables in the database, namely {\tt pgbench\_branches}, {\tt pgbench\_tellers}, {\tt pgbench\_accounts}, and {\tt pgbench\_history}.
\item \textbf{insert record}: inserts 15, 150 and 1,500,000 records into the aforementioned tables expect {\tt pgbench\_history} respectively.
\item \textbf{update and select record}: executes \emph{pgbench} built-in TPC-B-like transaction with prepared and ad-hoc queries:  updating records in table {\tt pgbench\_branches}, {\tt pgbench\_tellers},and {\tt pgbench\_accounts}, and then doing queries, finally inserting a record into table {\tt pgbench\_history}.
\end{itemize}
\subsubsection{Statistics}
During sandbox mining, the eight containers execute approximately 5,340,000 system calls. The number of system call execution of the eight containers is shown in Fig. \ref{fig:syscall_histgram}.  We can see that the number of system call execution goes to thousands or even millions. Thus tracing and analyzing system calls on a real-time environment will cause a considerate performance penalty. To achieve low performance penalty, we only trace and analyze system calls in sandbox mining phase.  A decomposition of the most frequent system calls of each container is shown in Fig. \ref{fig:syscall_decomposition}. The system call with the highest frequency is {\tt recvfrom()} which is used to receive a message from a socket. The corresponding system call {\tt sendto()} which is used to send a message on a socket has high frequency as well. The system calls that monitor multiple file descriptors are also prominent, such as {\tt epoll\_ctl()} and {\tt epoll\_wait()}. System calls that access filesystem are also executed frequently, such as {\tt read()} and {\tt write()}.

\pgfplotstableread {
container	number
Redis	1889805
MongoDB	1650448
PostgreSQL	880817
MySQL	410740
Python	373670
Apache	89765
Node.js	29539
Nginx	15743
}\syscallnum

\begin{figure}[t]
\footnotesize
\begin{tikzpicture}
  \begin{axis}[
      ybar,
      width=0.48\textwidth,
      height=.6\linewidth,
      legend style={at={(0.6,0.95)},
                        anchor=north,font=\footnotesize\selectfont},
      symbolic x coords={Redis,MongoDB,PostgreSQL,MySQL,Python,Apache,Node.js,Nginx},
      x tick label style={rotate=60,anchor=east},
      ymin=0,ymax=2000000,
      scaled ticks=false,
      yticklabel style={/pgf/number format/fixed},
      xtick=data,
      ymajorgrids=true,
      ]
    \addplot table[x=container,y=number]{\syscallnum};
  \end{axis}
\end{tikzpicture}\vspace{-0.2cm}
\caption{Number of system call execution of the containers.}\vspace{-0.5cm}
\label{fig:syscall_histgram}
\end{figure}
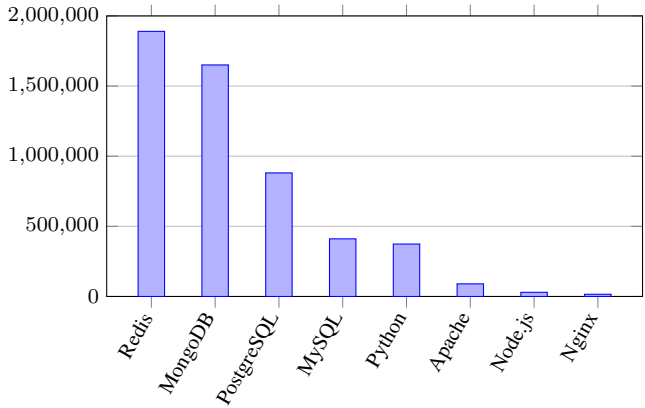
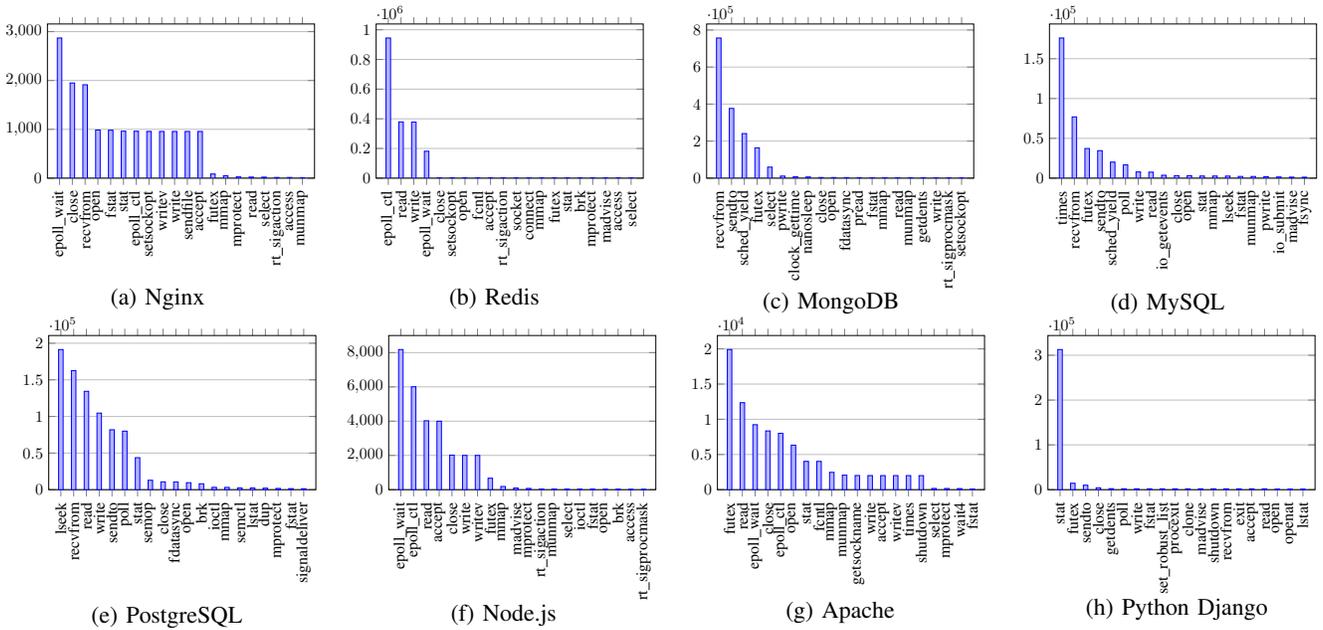
\begin{figure*}
    \begin{subfigure}[t]{0.24\textwidth}
        \centering
            \begin{tikzpicture}[scale=0.6, baseline]
            \begin{axis} [
                ybar,
                bar width=.1cm,
                width=7.5cm, height=5cm,
                ymajorgrids=true,
                xticklabels from table={nginx_syscall_freq.dat}{name},
                xtick=data,
                ymin=0,
                x tick label style={rotate=90,anchor=east},
                enlarge x limits=0.05,
                ]
\addplot table[y=frequency, x expr=\coordindex]{nginx_syscall_freq.dat};
\end{axis}
            \end{tikzpicture}
  \vfill
        \caption{Nginx}
    \end{subfigure}
    \begin{subfigure}[t]{0.24\textwidth}
        \centering
            \begin{tikzpicture}[scale=0.6, baseline]
            \begin{axis} [
                ybar,
                bar width=.1cm,
                width=7.5cm, height=5cm,
                ymajorgrids=true,
                xticklabels from table={redis_syscall_freq.dat}{name},
                xtick=data,
                ymin=0,
                x tick label style={rotate=90,anchor=east},
                enlarge x limits=0.05,
                ]
\addplot table[y=frequency, x expr=\coordindex]{redis_syscall_freq.dat};
\end{axis}
            \end{tikzpicture}
        \caption{Redis}
    \end{subfigure}
    \begin{subfigure}[t]{0.24\textwidth}
        \centering
            \begin{tikzpicture}[scale=0.6, baseline]
            \begin{axis} [
                ybar,
                bar width=.1cm,
                width=7.5cm, height=5cm,
                ymajorgrids=true,
                xticklabels from table={mongo_syscall_freq.dat}{name},
                xtick=data,
                ymin=0,
                x tick label style={rotate=90,anchor=east},
                enlarge x limits=0.05,
                ]
\addplot table[y=frequency, x expr=\coordindex]{mongo_syscall_freq.dat};
\end{axis}
            \end{tikzpicture}

\vspace{-0.6\baselineskip}
        \caption{MongoDB}
    \end{subfigure}
    \begin{subfigure}[t]{0.24\textwidth}
        \centering
            \begin{tikzpicture}[scale=0.6, baseline]
            \begin{axis} [
                ybar,
                bar width=.1cm,
                width=7.5cm, height=5cm,
                ymajorgrids=true,
                xticklabels from table={mysql_syscall_freq.dat}{name},
                xtick=data,
                ymin=0,
                x tick label style={rotate=90,anchor=east},
                enlarge x limits=0.05,
                ]
\addplot table[y=frequency, x expr=\coordindex]{mysql_syscall_freq.dat};
\end{axis}
            \end{tikzpicture}
        \caption{MySQL}
    \end{subfigure}
    \\
    \vspace{0.6\baselineskip}
    \begin{subfigure}[t]{0.24\textwidth}
        \centering
            \begin{tikzpicture}[scale=0.6, baseline]
            \begin{axis} [
                ybar,
                bar width=.1cm,
                width=7.5cm, height=5cm,
                ymajorgrids=true,
                xticklabels from table={postgres_syscall_freq.dat}{name},
                xtick=data,
                ymin=0,
                x tick label style={rotate=90,anchor=east},
                enlarge x limits=0.05,
                ]
\addplot table[y=frequency, x expr=\coordindex]{postgres_syscall_freq.dat};
\end{axis}
            \end{tikzpicture}
        \caption{PostgreSQL}
    \end{subfigure}
       \begin{subfigure}[t]{0.24\textwidth}
        \centering
            \begin{tikzpicture}[scale=0.6, baseline]
            \begin{axis} [
                ybar,
                bar width=.1cm,
                width=7.5cm, height=5cm,
                ymajorgrids=true,
                xticklabels from table={node_syscall_freq.dat}{name},
                xtick=data,
                ymin=0,
                x tick label style={rotate=90,anchor=east},
                enlarge x limits=0.05,
                ]
\addplot table[y=frequency, x expr=\coordindex]{node_syscall_freq.dat};
\end{axis}
            \end{tikzpicture}
\vspace{-0.6\baselineskip}
        \caption{Node.js}
    \end{subfigure}
       \begin{subfigure}[t]{0.24\textwidth}
        \centering
            \begin{tikzpicture}[scale=0.6, baseline]
            \begin{axis} [
                ybar,
                bar width=.1cm,
                width=7.5cm, height=5cm,
                ymajorgrids=true,
                xticklabels from table={httpd_syscall_freq.dat}{name},
                xtick=data,
                ymin=0,
                x tick label style={rotate=90,anchor=east},
                enlarge x limits=0.05,
                ]
\addplot table[y=frequency, x expr=\coordindex]{httpd_syscall_freq.dat};
\end{axis}
            \end{tikzpicture}
  \vspace{-0.2\baselineskip}
        \caption{Apache}
    \end{subfigure}
       \begin{subfigure}[t]{0.24\textwidth}
        \centering
            \begin{tikzpicture}[scale=0.6, baseline]
            \begin{axis} [
                ybar,
                bar width=.1cm,
                width=7.5cm, height=5cm,
                ymajorgrids=true,
                xticklabels from table={python_syscall_freq.dat}{name},
                xtick=data,
                ymin=0,
                x tick label style={rotate=90,anchor=east},
                enlarge x limits=0.05,
                ]
\addplot table[y=frequency, x expr=\coordindex]{python_syscall_freq.dat};
\end{axis}
            \end{tikzpicture}
 \vspace{-0.6\baselineskip}
        \caption{Python Django}
        \label{fig:python_syscall}
    \end{subfigure}\vspace{-0.3cm}
    \caption{Histogram of system call frequency for each of the containers.}\vspace{-0.3cm}
        \label{fig:syscall_decomposition}
\end{figure*}
\subsection{Growth of System Calls}
\label{sec:exp_syscall}
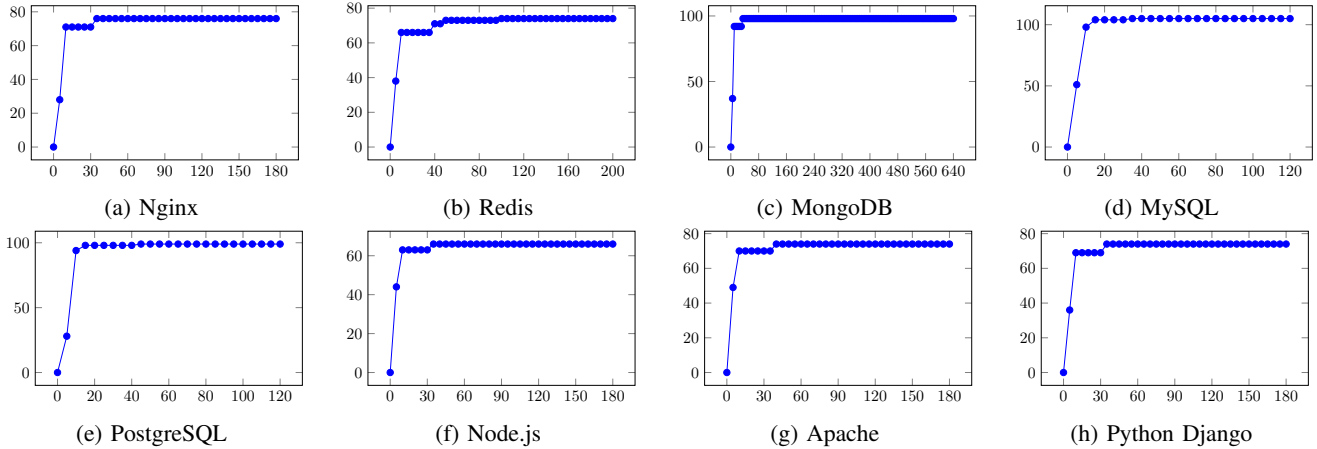
\begin{figure*}
    \begin{subfigure}[t]{0.24\textwidth}
        \centering
            \begin{tikzpicture}[scale=0.6, baseline]
            \begin{axis} [
                width=7.5cm, height=5cm,
                xtick={0,30,60,90,120,150,180}]
\addplot [mark=*,color=blue,mark size=2]
	table[y=SystemCall, x=Time]{nginx_syscall.dat};
\end{axis}
            \end{tikzpicture}
        \caption{Nginx}
        \label{fig:nginx_syscall}
    \end{subfigure}
    \begin{subfigure}[t]{0.24\textwidth}
        \centering
            \begin{tikzpicture}[scale=0.6, baseline]
            \begin{axis} [
                width=7.5cm, height=5cm,
                xtick={0,40,80,120,160,200}]
\addplot [mark=*,color=blue,mark size=2]
	table[y=SystemCall, x=Time]{redis_syscall.dat};
\end{axis}
            \end{tikzpicture}
        \caption{Redis}
        \label{fig:redis_syscall}
    \end{subfigure}
    \begin{subfigure}[t]{0.24\textwidth}
        \centering
            \begin{tikzpicture}[scale=0.6, baseline]
            \begin{axis} [
                width=7.5cm, height=5cm,
                xtick={0,80,160,240,320,400,480,560,640}]
\addplot [mark=*,color=blue,mark size=2]
	table[y=SystemCall, x=Time]{mongo_syscall.dat};
\end{axis}
            \end{tikzpicture}
        \caption{MongoDB}
        \label{fig:mongo_syscall}
    \end{subfigure}
    \begin{subfigure}[t]{0.24\textwidth}
        \centering
            \begin{tikzpicture}[scale=0.6, baseline]
            \begin{axis} [
                width=7.5cm, height=5cm,
                xtick={0,20,40,60,80,100,120}]
\addplot [mark=*,color=blue,mark size=2]
	table[y=SystemCall, x=Time]{mysql_syscall.dat};
\end{axis}
            \end{tikzpicture}
        \caption{MySQL}
        \label{fig:mysql_syscall}
    \end{subfigure}

    \begin{subfigure}[t]{0.24\textwidth}
        \centering
            \begin{tikzpicture}[scale=0.6, baseline]
            \begin{axis} [
                width=7.5cm, height=5cm,
                xtick={0,20,40,60,80,100,120}]
\addplot [mark=*,color=blue,mark size=2]
	table[y=SystemCall, x=Time]{postgres_syscall.dat};
\end{axis}
            \end{tikzpicture}
        \caption{PostgreSQL}
        \label{fig:postgres_syscall}
    \end{subfigure}
       \begin{subfigure}[t]{0.24\textwidth}
        \centering
            \begin{tikzpicture}[scale=0.6, baseline]
            \begin{axis} [
                width=7.5cm, height=5cm,
                xtick={0,30,60,90,120,150,180}]
\addplot [mark=*,color=blue,mark size=2]
	table[y=SystemCall, x=Time]{node_syscall.dat};
\end{axis}
            \end{tikzpicture}
        \caption{Node.js}
        \label{fig:node_syscall}
    \end{subfigure}
       \begin{subfigure}[t]{0.24\textwidth}
        \centering
            \begin{tikzpicture}[scale=0.6, baseline]
            \begin{axis} [
                width=7.5cm, height=5cm,
                xtick={0,30,60,90,120,150,180}]
\addplot [mark=*,color=blue,mark size=2]
	table[y=SystemCall, x=Time]{httpd_syscall.dat};
\end{axis}
            \end{tikzpicture}
        \caption{Apache}
        \label{fig:httpd_syscall}
    \end{subfigure}
       \begin{subfigure}[t]{0.24\textwidth}
        \centering
            \begin{tikzpicture}[scale=0.6, baseline]
            \begin{axis} [
                width=7.5cm, height=5cm,
                xtick={0,30,60,90,120,150,180}]
\addplot [mark=*,color=blue,mark size=2]
	table[y=SystemCall, x=Time]{python_syscall.dat};
\end{axis}
            \end{tikzpicture}
        \caption{Python Django}
        \label{fig:python_syscall}
    \end{subfigure}
    \caption{Per-container system call saturation for the containers in TABLE \ref{tab:subjects}. y axis is the number of accessed system calls, x axis is seconds spent.}\vspace{-0.5cm}
        \label{fig:syscall}
\end{figure*}

Fig. \ref{fig:syscall} shows the system call saturation charts for the eight containers. We can see that six charts ``flatten'' before one minute mark, and the remaining two before two minutes. Our approach has discovered 76, 74, 98, 105, 99, 66, 73, and 74 system calls accessed by \emph{Nginx}, \emph{Redis}, \emph{MongoDB}, \emph{MySQL}, \emph{PostgreSQL}, \emph{Node.js}, \emph{Apache}, and \emph{Python Django} containers respectively. The number of accessed system calls is far less than 300+ of the default Docker sandbox. The attack surface is significantly reduced.

During the warm-up phase, the number of system calls accessed by each of the containers grows rapidly. After the warm-up phase, for all of the Web servers except \emph{Apache}, the simple HTTP request causes a further increase and the number of system calls converges; for \emph{Apache} container, \emph{httperf} causes a small increase and the number of system calls shows no change later. For \emph{Redis} container, connecting to the container via {\tt docker exec} causes a first increase after the warm-up phase; and later \emph{redis-benchmark} triggers a small increase. For \emph{MongoDB}, \emph{MySQL} and \emph{PostgreSQL} containers, \emph{mongo-perf}, \emph{sysbench} and \emph{pgbench} cause a small increase after the warm-up phase.

The answer of \textbf{RQ1} is: our approach can mine the saturated set of system calls within two minutes. The mined sandboxes   reduce the attack surface.

\vspace{0.1cm}\noindent\fbox{%
    \parbox{\linewidth} {%
         \emph{Sandbox mining quickly saturates accessed system calls.}
    }%
}

\subsection{False Alarm}
\subsubsection{Use cases}
Our approach stops discovering new accessed system calls before the testing ends. However, does this mean that the most important functionality of a container is actually found? To answer this question, we carefully read the documentation of the containers and prepared \emph{use cases} which reflect containers' typical usages. TABLE \ref{tab:false} provides a full list of the use cases. We implemented all of these use cases as automated {\tt bash} test cases, allowing for easy assessment and replication.

After mining the sandbox for a given container, the central question for the evaluation is whether these use cases would be impacted by the sandbox, i.e., a benign system call would be denied during sandbox enforcing. To recognize the impact of sandbox, we set the default action of sandboxes to be {\tt SCMP\_ACT\_KILL} in the experiment. When the mined sandbox denies a system call, the process which accesses the system call will be killed, and \emph{auditd} \cite{auditd} will log a message of type {\tt SECCOMP} for the failed system call. Note that the default action of our mined sandboxes is {\tt SCMP\_ACT\_ERRNO} in production.

\subsubsection{Results}
The ``Messages in \emph{auditd}'' column in TABLE \ref{tab:false} summarizes the number of messages logged by \emph{auditd}. We can see that no message is logged by \emph{audid} for the 30 use cases. The number of false alarm is zero.

The answer of \textbf{RQ2} is: we did not find any impact from the mined sandboxes on the regular functionalities of the containers. Even automatic testing of a small workload is suitable to cover sufficient ``normal'' behaviors for the use cases in TABLE \ref{tab:false}.

 \vspace{0.2cm}\noindent\fbox{%
    \parbox{0.965\linewidth} {%
         \emph{Mined sandboxes require no further adjustment on use cases.}
    }%
}
\begin{table*}
  \centering
  \caption{Use cases. \emph{auditd} logs a message when a system call is denied by the sandbox.}
    \begin{tabular} {p{0.1\linewidth} p{0.16\linewidth} p{0.44\linewidth} >{\centering\arraybackslash}p{0.13\linewidth}}
    \rowcolor[rgb]{ .651,  .651,  .651} \textbf{Container} & \textbf{Use Case} & \textbf{Functions} & \textbf{Messages in \emph{auditd}} \\
  Nginx & Access static page & Access default page {\tt index.html}, {\tt 50x.html} & - \\
    \rowcolor[rgb]{ .851,  .851,  .851}       & Access non-existent page & Access non-existent page {\tt hello.html} & - \\
    Redis & SET command & Connect to Redis server, set key to hold the string value & - \\
    \rowcolor[rgb]{ .851,  .851,  .851}       & GET command & Connect to Redis server, get the value of key & - \\
          & INCR command & Connect to Redis server, increment the number stored at key by one & - \\
    \rowcolor[rgb]{ .851,  .851,  .851}       & LPUSH command & Connect to Redis server, insert all the specified values at the head of the list stored at key. & - \\
          & LPOP command & Connect to Redis server,  remove and returns the first element of the list stored at key & - \\
    \rowcolor[rgb]{ .851,  .851,  .851}       & SADD command & Connect to Redis server,  add the specified members to the set stored at key & - \\
          & SPOP command & Connect to Redis server, remove and return one or more random elements from the set value store at key & - \\
    \rowcolor[rgb]{ .851,  .851,  .851}       & LRANGE command & Connect to Redis server, return the specified elements of the list stored at key & - \\
          & MSET command & Connect to Redis server, replace multiple existing values with new values & - \\
    \rowcolor[rgb]{ .851,  .851,  .851} MongoDB & insert & Connect to mongod, use database {\tt test}, insert record {\tt \{image:"redis",count:"1"\}} into collection {\tt falsealarm}, exit  & - \\
          & save  & Connect to mongod, use database {\tt test}, update record in collection {\tt falsealarm}, exit  & - \\
    \rowcolor[rgb]{ .851,  .851,  .851}       & find  & Connect to mongod, use database {\tt test}, list all records  in collection {\tt falsealarm}, exit  & - \\
    MySQL & CREATE DATABASE & Connect to MySQL server, create database  {\tt test}, list all databases, exit & - \\
    \rowcolor[rgb]{ .851,  .851,  .851}       & CREATE TABLE & Connect to MySQL server, use database  {\tt test}, create table {\tt FalseAlarm}, insert record, exit & - \\
          & INSERT & Connect to MySQL server, use database  {\tt test}, insert record into table {\tt FalseAlarm}, exit & - \\
    \rowcolor[rgb]{ .851,  .851,  .851}       & UPDATE & Connect to MySQL server, use database  {\tt test}, update record, exit  & - \\
          & SELECT & Connect to MySQL server, use database  {\tt test}, list all records, exit  & - \\
    \rowcolor[rgb]{ .851,  .851,  .851} PostgreSQL & CREATE DATABASE & Connect to PostgreSQL server,  create database {\tt test}, list all databases, exit & - \\
          & CREATE TABLE & Connect to PostgreSQL server, connect to database {\tt test}, create table  {\tt FalseAlarm}, exit & - \\
    \rowcolor[rgb]{ .851,  .851,  .851}       & INSERT & Connect to PostgreSQL server, connect to database {\tt test}, insert record into table  {\tt FalseAlarm}, exit  & - \\
          & UPDATE & Connect to PostgreSQL server, connect to database {\tt test}, update record in table  {\tt FalseAlarm}, exit  & - \\
    \rowcolor[rgb]{ .851,  .851,  .851}       & SELECT & Connect to PostgreSQL server, connect to database {\tt test}, list all records in table  {\tt FalseAlarm}, exit  & - \\
    Node.js & Access existent URI & Access {\tt /} & - \\
    \rowcolor[rgb]{ .851,  .851,  .851}       & Access non-existent URI & Access non-existent URI {\tt /hello} & - \\
    Apache & Access static page & Access default page {\tt index.html} & - \\
    \rowcolor[rgb]{ .851,  .851,  .851}       & Access non-existent page & Access non-existent page {\tt hello.html} & - \\
    Python Django & Access existent URI & Access {\tt /} & - \\
    \rowcolor[rgb]{ .851,  .851,  .851}       & Access non-existent URI & Access non-existent URI {\tt /hello} & - \\
    \end{tabular}%
  \label{tab:false}\vspace{-0.4cm}%
\end{table*}%
\subsection{Performance Evaluation}
To analyze the performance characteristics of our approach, we run the eight containers in three environments: 1) natively without sandbox as a baseline, 2) with a sandbox mined by our approach, and 3) with the default Docker sandbox.
We measure the throughput of each container as an end-to-end performance metric. To minimize the impact of network, we run each of the containers using host networking via {\tt docker run --net=host}. We repeat each experiment 10 times with a less than 5\% standard deviation.

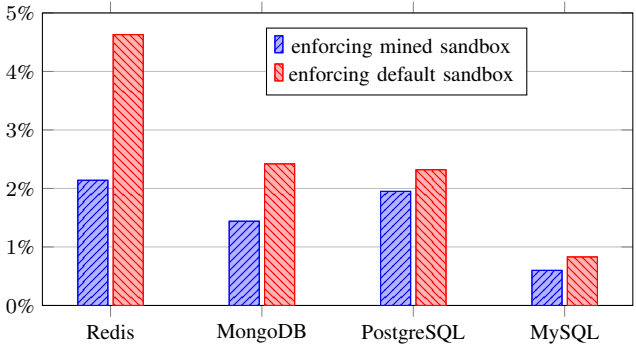
\begin{figure}[t]
\footnotesize
\begin{tikzpicture}
\footnotesize
  \begin{axis}[
      ybar,
      bar width=.4cm,
      width=0.52\textwidth,
      height=.618\linewidth,
      enlarge x limits=0.15,
      legend style={at={(0.6,0.95)},
                        anchor=north,font=\footnotesize\selectfont},
      symbolic x coords={Redis,MongoDB,PostgreSQL,MySQL},
      xtick=data,
      ytick={5,4,3,2,1,0},
      ymin=0,ymax=5,
      yticklabel={\pgfmathprintnumber\tick\%},
      ymajorgrids=true,
      ]
    \addplot
    [blue,fill=blue!30,
    pattern color=blue,
    postaction={
        pattern=north east lines
    }] table[x=container,y=customized]{\otherperfdata};
    \addplot
    [red,fill=red!30,
    pattern color=red,
    postaction={
        pattern=north west lines
    }]
    table[x=container, y=default]{\otherperfdata};
    \legend{enforcing mined sandbox, enforcing default sandbox}
  \end{axis}
\end{tikzpicture}
\caption{Percentage reduction of transactions per second (TPS) due to sandboxing.}\vspace{-0.6cm}
\label{fig:perf_other}
\end{figure}

For \emph{Redis}, \emph{MongoDB}, \emph{PostgreSQL} and \emph{MySQL} containers, we evaluate the \emph{transactions per second}  (TPS) of each container by running the aforementioned tools in Section \ref{sec:exp_syscall}.  The percentage reduction of TPS per container for \emph{Redis}, \emph{MongoDB}, \emph{PostgreSQL} and \emph{MySQL} is presented in Fig. \ref{fig:perf_other}. We notice that enforcing mined sandboxes incurs a small TPS reduction (0.6\% - 2.14\%) for the four containers. Mined sandboxes produce a slightly smaller TPS reduction than that of the default sandbox (0.83\% - 4.63\%). The reason is that the default sandbox contains more rules than mined sandboxes, and thus the corresponding \emph{BFP program} needs more computation during sandboxing.

\begin{figure*}[t]
\footnotesize
    \begin{subfigure}[b]{0.48\textwidth}
        \centering
            \begin{tikzpicture}[scale=0.9, baseline]
                \begin{axis} [
  width=8cm, height=5.5cm,
        xmin=0,xmax=10000,
        ymin=0,ymax=9000,
        ymajorgrids=true,
        scaled x ticks=false,
	]
	\addplot[mark=none,color=blue,line width=1.2pt] table[x=RequestRate, y=NONE]{perf_nginx.dat};
	\addplot[mark=none,color=red,line width=1.2pt,dashed] table[x=RequestRate, y=CUSTOMIZED]{perf_nginx.dat};
	\addplot[mark=none,color=gray,line width=1.2pt,densely dotted] table[x=RequestRate, y=DEFAULT]{perf_nginx.dat};
	\end{axis}
\end{tikzpicture}
        \caption{Nginx}
    \end{subfigure}
    \begin{subfigure}[b]{0.48\textwidth}
        \centering
            \begin{tikzpicture}[scale=0.9, baseline]
                            \begin{axis} [
  width=8cm, height=5.5cm,
        xmin=0,xmax=4000,
        ymin=0,ymax=3500,
        ymajorgrids=true,
        scaled x ticks=false,
	]
	\addplot[mark=none,color=blue,line width=1.2pt] table[x=RequestRate, y=NONE]{perf_node.dat};
	\addplot[mark=none,color=red,line width=1.2pt,dashed] table[x=RequestRate, y=CUSTOMIZED]{perf_node.dat};
	\addplot[mark=none,color=gray,line width=1.2pt,densely dotted] table[x=RequestRate, y=DEFAULT]{perf_node.dat};
	\end{axis}
            \end{tikzpicture}
        \caption{Node.js}
    \end{subfigure}
\\\\
    \begin{subfigure}[b]{0.48\textwidth}
        \centering
            \begin{tikzpicture}[scale=0.9, baseline]

                            \begin{axis} [
  width=8cm, height=5.5cm,
        xmin=0,xmax=4000,
        ymin=0,ymax=3500,
        ymajorgrids=true,
        scaled x ticks=false,
	]
	\addplot[mark=none,color=blue,line width=1.2pt] table[x=RequestRate, y=NONE]{perf_httpd.dat};
	\addplot[mark=none,color=red,line width=1.2pt,dashed] table[x=RequestRate, y=CUSTOMIZED]{perf_httpd.dat};
	\addplot[mark=none,color=gray,line width=1.2pt,densely dotted] table[x=RequestRate, y=DEFAULT]{perf_httpd.dat};
	\end{axis}
            \end{tikzpicture}
        \caption{Apache}
    \end{subfigure}
    \begin{subfigure}[b]{0.48\textwidth}
        \centering
            \begin{tikzpicture}[scale=0.9, baseline]
                            \begin{axis} [
  width=8cm, height=5.5cm,
        xmin=0,xmax=500,
        ymin=0,ymax=350,
        ymajorgrids=true,
        scaled x ticks=false,
	]
	\addplot[mark=none,color=blue,line width=1.2pt] table[x=RequestRate, y=NONE]{perf_python.dat};
	\addplot[mark=none,color=red,line width=1.2pt,dashed] table[x=RequestRate, y=CUSTOMIZED]{perf_python.dat};
	\addplot[mark=none,color=gray,line width=1.2pt,densely dotted] table[x=RequestRate, y=DEFAULT]{perf_python.dat};
	\end{axis}
            \end{tikzpicture}
        \caption{Python Django}
    \end{subfigure}
    \begin{center}
       \begin{tikzpicture}
        \begin{customlegend}[legend columns=4,legend style={align=left,draw=none,column sep=2ex},legend entries={without sandbox, with mined sandbox, with default sandbox}]
        \addlegendimage{mark=none,solid,color=blue,line width=1.2pt}
        \addlegendimage{mark=none,dashed,color=red,line width=1.2pt}
        \addlegendimage{mark=none,densely dotted,color=gray,line width=1.2pt}
        \end{customlegend}
     \end{tikzpicture}\vspace{-0.5cm}
    \end{center}
    \caption{Comparison of per-container reply rate for Nginx, Node.js, Apache, and Python Django without sandbox, with sandbox mined by our approach, and with default sandbox. y axis is response rate (responses per second), x axis is request rate (requests per second).}\vspace{-0.5cm}
\label{fig:perf_webserver}
\end{figure*}
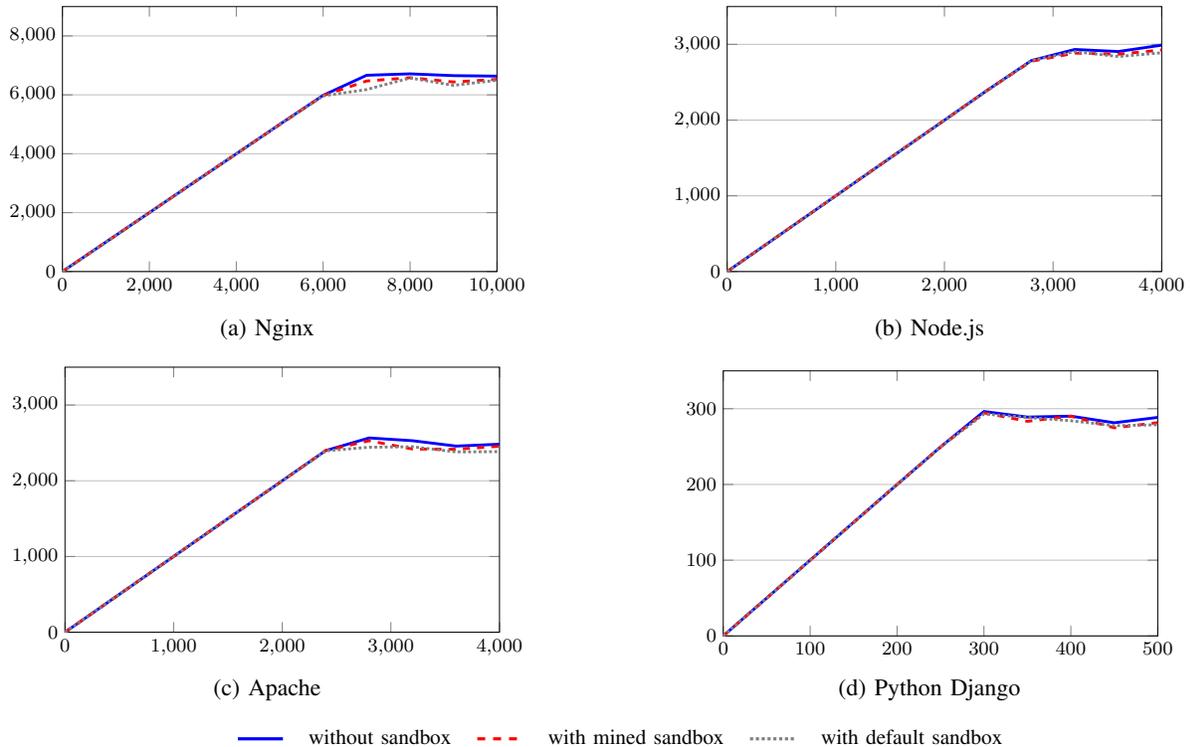

For Web server containers, we evaluate the throughput, i.e., \emph{responses per second}, of each container by running \emph{httperf} tool. To measure the response rate of each container, we increase the number of requests per second that are sent to the container. The result is shown in Fig. \ref{fig:perf_webserver}. Web server containers running with sandboxes achieve a performance very similar to that of the containers running without sandboxes. We can see that the achieved throughput increases linearly with offered load until the container starts to become saturated. The saturation points of \emph{Nginx}, \emph{Node.js}, \emph{Apache} and \emph{Python Django} are around 7,000, 3,000, 2,500 and 300 requests per second respectively. After offered load is increased beyond that point, the response rate of the container starts to fall off slightly.

The answer of \textbf{RQ3} is: enforcing system call policies adds overhead to a container's end-to-end performance, but the overall increase is small.

\vspace{0.1cm}\noindent\fbox{%
    \parbox{0.965\linewidth} {%
         \emph{Sandboxes incur a small end-to-end performance overhead.}
    }%
}
\pgfplotstableread {
container	customized	default
Redis	2.14	4.63
MongoDB	1.44	2.42
PostgreSQL	1.95	2.32
MySQL	0.60	0.83
}\otherperfdata


\section{Threats and Limitations}
\label{sec:threat}
System call access is either benign or malicious. Our approach automatically decides on whether a system call accessed by a container should be allowed. As we do not assume a specification of what makes a benign or malicious system call access for a container, we face two risks:
\begin{itemize}
\item \textbf{False positive.} A \emph{false positive} occurs when a benign system call is mistakenly prohibited by the sandbox, degrading a container's functionality. In our setting, a false alarm happens if some benign system call is not seen during mining phase, and thus not added to sandbox rules to be allowed. The number of false alarms can be reduced by better testing.
\item\textbf{False negative.} A \emph{false negative} occurs when a malicious system call is mistakenly allowed by the sandbox. In our setting, a false alarm can happen in two ways:
    \begin{itemize}
    \item \textbf{False negative allowed during sandbox enforcing.} The inferred sandbox rules may be too coarse, and thus allow future malicious system calls. For instance, a container may access system calls {\tt mmap()}, {\tt mprotect()} and {\tt munmap()} as benign behaviors. However, \emph{code injection} attack could also invoke these system calls to change memory protection. This issue can be addressed by inferring more fine-grained sandbox rules.
    \item \textbf{False negative seen during sandbox mining.} The container may be initially malicious. We risk to mine the malicious behaviors of the container during mining phase. Thus malicious system calls would be included in the sandbox rules. This issue can be addressed by identifying malicious behaviors during mining phase.
    \end{itemize}
\end{itemize}

 Although our experimental results demonstrate the feasibility of sandbox mining for containers, our sample of containers is small and the containers are database systems and Web servers. For other containers, we have to design different testing. In addition, some containers may comprise multiple processes which have distinct responsibilities, for instance, a Linux, Apache, MySQL and PHP (LAMP) stack in one container. This may increase attack surface, and lead to more false negatives.

The set of use cases we have prepared for assessing the risk of false alarms (TABLE \ref{tab:false}) does not and cannot cover the entire range of functionalities of the analyzed containers. Although we assume that the listed user cases represent the most important functionalities, other usage may yield different results.

Finally, in the absence of a specification, a mined policy cannot express whether a system call is benign or malicious. Although our approach cannot eliminate the risks of false positives and false negatives, we do reduce the attack surface by detecting and preventing unexpected behavior.


\section{Conclusion and Future Work}
\label{sec:conclusion}
In this paper, we present an approach to mine sandboxes for Linux containers. The approach explores the behaviors of a container by automatically running test suites. From the execution trace, the approach extracts set of system calls accessed by the container during the mining phase, and translates the system calls into sandbox rules. During sandbox enforcement, the mined sandbox confines the container by restricting its access to system calls. Our evaluation shows that our approach can efficiently mine sandboxes for containers and substantially reduce the attack surface. In our experiment, automatic testing sufficiently covers container behaviors and sandbox enforcement incurs low overhead.

In the future, we would like to mine more fine-grained sandbox policy, taking into account the system call arguments, temporal features of system calls, internal states of a container, or data flow from and to sensitive resources. More Fine-grained sandbox may lead to more false positives and increase performance overhead. We have to search for sweet spots that both minimize false positives and performance overhead. Meanwhile, we have to avoid \emph{Time-of-check-to-time-of-use} (TOCTTOU) problems when examining system call arguments.
We also plan to leverage modern test case generation techniques to systematically explore container behaviors. This may help to cover more normal behaviors of a container.
In addition, for now we enforce one system call policy on a whole container. Whereas a container may comprise multiple processes which have distinct behaviors. To further reduce the attack surface, We could enforce a distinct policy for each process which corresponds to the behavior of that process.

\section*{Acknowledgment}
We thank the anonymous reviewers for their insightful comments. This research is supported by NSFC Program (No. 61602403), and National Key Technology R\&D Program of the Ministry of Science and Technology of China (No. 2015BAH17F01).

\balance

\bibliographystyle{IEEEtran}
\bibliography{seccomp_docker}

\begin{thebibliography}{10}
\providecommand{\url}[1]{#1}
\csname url@samestyle\endcsname
\providecommand{\newblock}{\relax}
\providecommand{\bibinfo}[2]{#2}
\providecommand{\BIBentrySTDinterwordspacing}{\spaceskip=0pt\relax}
\providecommand{\BIBentryALTinterwordstretchfactor}{4}
\providecommand{\BIBentryALTinterwordspacing}{\spaceskip=\fontdimen2\font plus
\BIBentryALTinterwordstretchfactor\fontdimen3\font minus
  \fontdimen4\font\relax}
\providecommand{\BIBforeignlanguage}[2]{{%
\expandafter\ifx\csname l@#1\endcsname\relax
\typeout{** WARNING: IEEEtran.bst: No hyphenation pattern has been}%
\typeout{** loaded for the language `#1'. Using the pattern for}%
\typeout{** the default language instead.}%
\else
\language=\csname l@#1\endcsname
\fi
#2}}
\providecommand{\BIBdecl}{\relax}
\BIBdecl

\bibitem{paas}
G.~I.~A. Inc., ``{Platform as a Service PaaS Market Trends},''
  \url{http://www.strategyr.com/MarketResearch/Platform_as_a_Service_PaaS_Market_Trends.asp},
  2015, [Online; accessed 2016-08-16].

\bibitem{merkel2014docker}
D.~Merkel, ``Docker: lightweight linux containers for consistent development
  and deployment,'' \emph{Linux Journal}, vol. 2014, no. 239, p.~2, 2014.

\bibitem{felter2015updated}
W.~Felter, A.~Ferreira, R.~Rajamony, and J.~Rubio, ``An updated performance
  comparison of virtual machines and linux containers,'' in \emph{Proceedings
  of the 2015 IEEE International Symposium on Performance Analysis of Systems
  and Software (ISPASS 2015)}.\hskip 1em plus 0.5em minus 0.4em\relax IEEE,
  2015, pp. 171--172.

\bibitem{cve}
``{CVE-2016-0728},''
  \url{http://www.cve.mitre.org/cgi-bin/cvename.cgi?name=2016-0728}, [Online;
  accessed 2016-08-16].

\bibitem{garfinkel2003traps}
T.~Garfinkel \emph{et~al.}, ``Traps and pitfalls: Practical problems in system
  call interposition based security tools,'' in \emph{Network and Distributed
  System Security Symposium (NDSS 2003)}, vol.~3, 2003, pp. 163--176.

\bibitem{goldberg1996secure}
I.~Goldberg, D.~Wagner, R.~Thomas, E.~A. Brewer \emph{et~al.}, ``A secure
  environment for untrusted helper applications: Confining the wily hacker,''
  in \emph{Proceedings of the Conference on USENIX Security Symposium}, 1996.

\bibitem{provos2003improving}
N.~Provos, ``Improving host security with system call policies,'' in
  \emph{Proceedings of the Conference on USENIX Security Symposium}, 2003.

\bibitem{acharya2000mapbox}
A.~Acharya and M.~Raje, ``Mapbox: Using parameterized behavior classes to
  confine untrusted applications,'' in \emph{Proceedings of the conference on
  USENIX Security Symposium}.\hskip 1em plus 0.5em minus 0.4em\relax USENIX
  Association, 2000.

\bibitem{fraser1999hardening}
T.~Fraser, L.~Badger, and M.~Feldman, ``Hardening cots software with generic
  software wrappers,'' in \emph{Proceedings 1999 IEEE Symposium on Security and
  Privacy (S\&P 1999)}.\hskip 1em plus 0.5em minus 0.4em\relax IEEE, 1999, pp.
  2--16.

\bibitem{ko2000detecting}
C.~Ko, T.~Fraser, L.~Badger, and D.~Kilpatrickv, ``Detecting and countering
  system intrusions using software wrappers,'' in \emph{Proceedings of the
  Conference on USENIX Security Symposium}, 2000, pp. 1157--1168.

\bibitem{kim2013practical}
T.~Kim and N.~Zeldovich, ``Practical and effective sandboxing for non-root
  users,'' in \emph{Proceedings of the Conference on USENIX Annual Technical
  Conference (USENIX ATC 13)}, 2013, pp. 139--144.

\bibitem{jamrozik2016mining}
K.~Jamrozik, P.~von Styp-Rekowsky, and A.~Zeller, ``Mining sandboxes,'' in
  \emph{Proceedings of the 38th International Conference on Software
  Engineering (ICSE 2016)}.\hskip 1em plus 0.5em minus 0.4em\relax ACM, 2016,
  pp. 37--48.

\bibitem{dockerseccomp}
``{Seccomp security profiles for Docker},''
  \url{https://docs.docker.com/engine/security/seccomp}, [Online; accessed
  2016-08-16].

\bibitem{garfinkel2004ostia}
T.~Garfinkel, B.~Pfaff, M.~Rosenblum \emph{et~al.}, ``Ostia: A delegating
  architecture for secure system call interposition,'' in \emph{Network and
  Distributed System Security Symposium (NDSS 2004)}, 2004.

\bibitem{wagner1999janus}
D.~A. Wagner, ``Janus: an approach for confinement of untrusted applications,''
  Ph.D. dissertation, Department of Electrical Engineering and Computer
  Sciences, University of California at Berkeley, 1999.

\bibitem{jain2000user}
K.~Jain and R.~Sekar, ``User-level infrastructure for system call
  interposition: A platform for intrusion detection and confinement,'' in
  \emph{Network and Distributed System Security Symposium (NDSS 2000)}, 2000.

\bibitem{hofmeyr1998intrusion}
S.~A. Hofmeyr, S.~Forrest, and A.~Somayaji, ``Intrusion detection using
  sequences of system calls,'' \emph{Journal of computer security}, vol.~6,
  no.~3, pp. 151--180, 1998.

\bibitem{forrest1996sense}
S.~Forrest, S.~A. Hofmeyr, A.~Somayaji, and T.~A. Longstaff, ``A sense of self
  for unix processes,'' in \emph{Proceedings 1996 IEEE Symposium on Security
  and Privacy (S\&P 1996)}.\hskip 1em plus 0.5em minus 0.4em\relax IEEE, 1996,
  pp. 120--128.

\bibitem{wagner2001intrusion}
D.~Wagner and R.~Dean, ``Intrusion detection via static analysis,'' in
  \emph{Proceedings 2001 IEEE Symposium on Security and Privacy (S\&P
  2001)}.\hskip 1em plus 0.5em minus 0.4em\relax IEEE, 2001, pp. 156--168.

\bibitem{bhatkar2006dataflow}
S.~Bhatkar, A.~Chaturvedi, and R.~Sekar, ``Dataflow anomaly detection,'' in
  \emph{Proceedings 2006 IEEE Symposium on Security and Privacy (S\&P
  2006)}.\hskip 1em plus 0.5em minus 0.4em\relax IEEE, 2006, pp. 15--pp.

\bibitem{kiriansky2002secure}
V.~Kiriansky, D.~Bruening, S.~P. Amarasinghe \emph{et~al.}, ``Secure execution
  via program shepherding,'' in \emph{Proceedings of the Conference on USENIX
  Security Symposium}, vol.~92, 2002, p.~84.

\bibitem{warrender1999detecting}
C.~Warrender, S.~Forrest, and B.~Pearlmutter, ``Detecting intrusions using
  system calls: Alternative data models,'' in \emph{Proceedings 1999 IEEE
  Symposium on Security and Privacy (S\&P 1999)}.\hskip 1em plus 0.5em minus
  0.4em\relax IEEE, 1999, pp. 133--145.

\bibitem{somayaji2000automated}
A.~Somayaji and S.~Forrest, ``Automated response using system-call delay,'' in
  \emph{Proceedings of the Conference on USENIX Security Symposium}, 2000, pp.
  185--197.

\bibitem{sekar2001fast}
R.~Sekar, M.~Bendre, D.~Dhurjati, and P.~Bollineni, ``A fast automaton-based
  method for detecting anomalous program behaviors,'' in \emph{Proceedings 2001
  IEEE Symposium on Security and Privacy (S\&P 2001)}.\hskip 1em plus 0.5em
  minus 0.4em\relax IEEE, 2001, pp. 144--155.

\bibitem{mutz2006anomalous}
D.~Mutz, F.~Valeur, G.~Vigna, and C.~Kruegel, ``Anomalous system call
  detection,'' \emph{ACM Transactions on Information and System Security
  (TISSEC)}, vol.~9, no.~1, pp. 61--93, 2006.

\bibitem{seccompbpf}
``{Yet another new approach to seccomp},''
  \url{http://lwn.net/Articles/475043}, [Online; accessed 2016-08-16].

\bibitem{seccomp}
``{Seccomp and sandboxing},'' \url{http://lwn.net/Articles/475043}, [Online;
  accessed 2016-08-16].

\bibitem{json}
``{JSON},'' \url{http://www.json.org}, [Online; accessed 2016-08-16].

\bibitem{saltzer1975protection}
J.~H. Saltzer and M.~D. Schroeder, ``The protection of information in computer
  systems,'' \emph{Proceedings of the IEEE}, vol.~63, no.~9, pp. 1278--1308,
  1975.

\bibitem{kruegel2003detection}
C.~Kruegel, D.~Mutz, F.~Valeur, and G.~Vigna, ``On the detection of anomalous
  system call arguments,'' in \emph{European Symposium on Research in Computer
  Security (ESORICS 2003)}.\hskip 1em plus 0.5em minus 0.4em\relax Springer,
  2003, pp. 326--343.

\bibitem{fetzer2008switchblade}
C.~Fetzer and M.~S{\"u}{\ss}kraut, ``Switchblade: enforcing dynamic
  personalized system call models,'' in \emph{ACM SIGOPS Operating Systems
  Review}, vol.~42, no.~4.\hskip 1em plus 0.5em minus 0.4em\relax ACM, 2008,
  pp. 273--286.

\bibitem{gao2006behavioral}
D.~Gao, M.~K. Reiter, and D.~Song, ``Behavioral distance measurement using
  hidden markov models,'' in \emph{International Workshop on Recent Advances in
  Intrusion Detection}.\hskip 1em plus 0.5em minus 0.4em\relax Springer, 2006,
  pp. 19--40.

\bibitem{endler1998intrusion}
D.~Endler, ``Intrusion detection. applying machine learning to solaris audit
  data,'' in \emph{Proceedings of the 14th Annual Computer Security
  Applications Conference (ACSAC 1998)}.\hskip 1em plus 0.5em minus 0.4em\relax
  IEEE, 1998, pp. 268--279.

\bibitem{liao2002use}
Y.~Liao and V.~R. Vemuri, ``Use of k-nearest neighbor classifier for intrusion
  detection,'' \emph{Computers \& Security}, vol.~21, no.~5, pp. 439--448,
  2002.

\bibitem{zeller2015test}
A.~Zeller, ``Test complement exclusion: Guarantees from dynamic analysis,'' in
  \emph{Proceedings of the 2015 IEEE 23rd International Conference on Program
  Comprehension (ICPC 2015)}.\hskip 1em plus 0.5em minus 0.4em\relax IEEE
  Press, 2015, pp. 1--2.

\bibitem{whalen2001introduction}
S.~Whalen, ``An introduction to arp spoofing,'' \emph{Node99 [Online Document],
  April}, 2001.

\bibitem{sysdig}
``{sysdig},'' \url{http://www.sysdig.org}, [Online; accessed 2016-08-16].

\bibitem{helloworld}
``{hello-world},'' \url{https://hub.docker.com/_/hello-world}, [Online;
  accessed 2016-08-16].

\bibitem{runc_standard_init_linux_v011}
``{runc libcontainer version 0.1.1},''
  \url{https://github.com/opencontainers/runc/blob/v0.1.1/libcontainer/standard_init_linux.go},
  [Online; accessed 2016-08-16].

\bibitem{ptrace}
``{Ptrace documentation},'' \url{https://lwn.net/Articles/446593}, [Online;
  accessed 2016-08-16].

\bibitem{dockerhub}
``{Docker Hub},'' \url{https://hub.docker.com/explore}, [Online; accessed
  2016-08-16].

\bibitem{django}
``{Django: a high-level Python Web framework},''
  \url{https://www.djangoproject.com}, [Online; accessed 2016-08-16].

\bibitem{mosberger1998httperf}
D.~Mosberger and T.~Jin, ``httperf: a tool for measuring web server
  performance,'' \emph{ACM SIGMETRICS Performance Evaluation Review}, vol.~26,
  no.~3, pp. 31--37, 1998.

\bibitem{redisbench}
``{How fast is Redis?}'' \url{http://redis.io/topics/benchmarks}, [Online;
  accessed 2016-08-16].

\bibitem{mongoperf}
``{Mongo-perf},'' \url{https://github.com/mongodb/mongo-perf}, [Online;
  accessed 2016-08-16].

\bibitem{sysbench}
``{SysBench},'' \url{https://github.com/akopytov/sysbench}, [Online; accessed
  2016-08-16].

\bibitem{pgbench}
``{pgbench},'' \url{https://www.postgresql.org/docs/9.3/static/pgbench.html},
  [Online; accessed 2016-08-16].

\bibitem{auditd}
``{auditd},'' \url{http://linux.die.net/man/8/auditd}, [Online; accessed
  2016-08-16].

\end{thebibliography}
\end{document}